\documentclass[journal,10pt]{IEEEtran}

\usepackage{graphicx}
\usepackage{latexsym}
\usepackage{times}
\usepackage{amsmath,amssymb,amsthm}

\usepackage{cite}

\usepackage{subfigure}

\usepackage{lipsum}
\usepackage{multirow}

\usepackage{mathtools}
\DeclarePairedDelimiter\abs{\lvert}{\rvert}%
\DeclarePairedDelimiter\norm{\lVert}{\rVert}%

\makeatletter
\let\oldabs\abs
\def\abs{\@ifstar{\oldabs}{\oldabs*}}
\let\oldnorm\norm
\def\norm{\@ifstar{\oldnorm}{\oldnorm*}}
\makeatother

\usepackage{color,xcolor,colortbl}

\usepackage{algorithm2e} 
\makeatletter
\renewcommand{\@algocf@capt@plain}{above}
\makeatother

\usepackage[utf8]{inputenc} 

\makeatletter
\def\markboth#1#2{\def\leftmark{\@IEEEcompsoconly{\sffamily}\MakeUppercase{\protect#1}}%
\def\rightmark{\@IEEEcompsoconly{\sffamily}\MakeUppercase{\protect#2}}}
\makeatother

\usepackage[english]{babel} 
\usepackage{url}

\newcommand{\IGNORE}[1]{}

\newcommand{\firstAuthor}{Ricardo J. Rodr\'{i}guez}
\newcommand{\firstAuthorShort}{R.~J.~Rodr\'{i}guez}
\newcommand{\DIIS}{Dpto. de Informática e Ingeniería de Sistemas}
\newcommand{\UZ}{Universidad de Zaragoza, Calle Mar\'{i}a de Luna 1, 50018
Zaragoza, Spain}
\newcommand{\mailRJ}{rjrodriguez@unizar.es}
\newcommand{\mailTo}[1]{\url{#1}}

\newcommand{\thirdAuthor}{Julián Fernández-Navajas}
\newcommand{\thirdAuthorShort}{J. Fernández-Navajas}
\newcommand{\secondAuthor}{Jos\'{e} Luis Salazar}
\newcommand{\secondAuthorShort}{J.~L.~Salazar}
\newcommand{\DIEC}{Dpto. de Ingeniería Electrónica y Comunicaciones}
\newcommand{\mailJulian}{navajas@unizar.es}
\newcommand{\mailSalazar}{jsalazar@unizar.es}

\newcommand{\ackText}{$^*$Corresponding author. This work was supported in part by the 
Aragonese Government under {\em Programa de Proyectos Estratégicos de Grupos de Investigación} (
T21-20R and T31-20R) and by the Spanish Ministry of Economy, Industry and Competitiveness (TIN2015-64770-R).}

\newcommand{\paperTitle}{{\em Sittin'On the Dock of the (WiFi) Bay}:\\On the Frame Aggregation under IEEE 802.11 DCF}

\newcommand{\authorNames}{{\firstAuthor}, {\secondAuthor}, {\thirdAuthor}}
\newcommand{\authorNamesShort}{{\firstAuthorShort}, {\secondAuthorShort}, {\thirdAuthorShort}}

\newcommand{\keywordsList}{Wireless communication, Multiplexing}

\usepackage{lipsum}
\usepackage{balance}
\usepackage{hyperref}
\usepackage{breakurl}

\hypersetup{
  pdftitle={{\paperTitle} - {\authorNamesShort}},%
  pdfauthor={\authorNames},%
  pdfproducer={\LaTeX},%
  pdfsubject={{\paperTitle} - {\authorNamesShort}},%
  pdfkeywords={\keywordsList}%
}

\usepackage{tikz}
\usetikzlibrary{arrows}
\usetikzlibrary{shapes}
\usetikzlibrary{patterns}
\usetikzlibrary{positioning}
\usetikzlibrary{shapes.multipart}
\usetikzlibrary{decorations.pathreplacing}
\usetikzlibrary{calc}
\usetikzlibrary{fit}

\usepackage{breqn}

\begin{document}

\title{\paperTitle}

\author{{\firstAuthor}$^*$\IEEEmembership{,~Member,~IEEE}, {\secondAuthor}, and~{\thirdAuthor}
\thanks{\ackText}
\thanks{{\firstAuthor} is with {\DIIS}, {\UZ}. E-mail: {\mailRJ}}
\thanks{{\secondAuthor} and {\thirdAuthor} are with {\DIEC}, {\UZ}. E-mail: {\mailJulian}, {\mailSalazar}}
}


\maketitle

\begin{abstract}
It is well known that frame aggregation in Internet communications improves  transmission efficiency. However, it also causes a delay that for some real-time communications is inappropriate, thus creating a trade-off  between efficiency and delay. In this paper, we establish the conditions for frame aggregation under the IEEE 802.11 DCF  protocol to be beneficial on average delay. To do so, we first describe the transmission time in IEEE 802.11 in a stochastic framework and then we calculate the optimal value of the frames that, when aggregated, saves transmission time in the long term. Our findings, discussed with numerical experimentation, show that frame aggregation reduces transmission congestion and transmission delays.
\end{abstract}


\section{Introduction}

WiFi protocols have evolved to achieve higher and higher transmission speeds, giving rise to different standards, such as 802.11a/b/g/n/ac. Older standards are superseded by newer standards and must coexist in the same implementation to ensure backward compatibility. Those with a lower rate provide greater coverage, while those with a higher rate force the terminals to be closer. When the terminals are far away, there is a lower transmission rate, so the aggregation of frames in WiFi has been seen as an improvement in efficiency (less bandwidth overhead), but it increases the risk of causing delays. For instance, this situation can be problematic in real-time services such as VoIP, video conferencing, or online gaming, where the transmission delay must not exceed certain limit values to satisfy the user experience. In this paper, we study how frame aggregation causes delays and how this deterioration can be quantified,  especially for those standards (or lower rate situations) where it is more relevant to give a correct solution (e.g., 802.11b/g).

IEEE 802.11 DCF (Distributed Coordination Function) is the medium access control used by WiFi~\cite{802.11-standard-07}, which is based on Carrier Sense Multiple Access/Collision Avoidance and operates at layer 2 of the OSI model. CSMA/CA is a network arbitration protocol which regulates communication between multiple nodes communicating through a single channel. 

CSMA/CA works as follows. When a node wants to transmit, it first checks if the channel is free. When the node detects that the medium is continuously free for a time defined by the DCF Interframe Space (DIFS), then it is allowed to transmit its data frame. The receiving node will send an acknowledgment (ACK) frame after a time, specified by SIFS (Short Interframe Space),  when the sent frame is successfully received. Conversely, when the node detects that the medium is busy, it postpones its transmission until the end of the current transmission. After an additional DIFS interval, the node generates a random backoff period for an additional deferral time before transmitting. That is, the node generates an integer $i$ uniformly distributed in the interval $\{0, 1, ..., CW\}$, where $CW$ is an integer within the range of  $aCWmin$ and $aCWmax$, i.e., $aCWmin \leq CW \leq aCWmax$. The  $aCWmin$ and $aCWmax$ values are the minimum and maximum time for the content window and depend on the characteristics of the physical layer. Then, the node waits for the $i$ $SLOT$ intervals before transmitting. The $SLOT$ value also depends on the characteristics of the physical layer.

When another node transmits before the backoff period ends, the countdown stops, and the remaining time is used on the next transmission attempt. When two nodes have the same backoff period  (or the remaining backoff) values, they will transmit at the same time and therefore their transmissions collide. The station detects the collision as it will not receive the ACK frame from the receiving node. When this happens, an exponential backoff algorithm is applied, i.e., $CW$ is doubled up to $aCWmax$ for the next transmission attempt.

In~\cite{D-PIMRC-08} a complete study of IEEE 802.11 DCF performance is presented  considering throughput, fairness, and delay. As the authors indicate, CSMA/CA supports long-term fairness by achieving the best performance for a small number of nodes, although it suffers short-term unfairness when the number of nodes exceeds the optimal value, as nodes whose transmission collides will increase their content window and then they will be less likely to access the medium.

Note that CSMA/CA exhibits some randomness as a random integer is computed to calculate the backoff period and the data payload of each frame (although it is limited to a maximum of 1500 bytes, it is likely to be different in each frame).  In this work, we want to show that under certain conditions, frame aggregation can reduce the average delay. In this regard, we first consider this randomness to better characterize the  transmission time of a frame under IEEE 802.11, which allows us to obtain an analytical solution to our problem. Then, we calculate the optimal value of the frames that, when aggregated, saves transmission time in the long term. Finally, we present and discuss some numerical experiments. 

This paper is organized as follows. Section~\ref{sec:related_work} reviews related work. Section~\ref{sec:stochastic} details the characterization of frame transmission time by means of random variables. Then, Section~\ref{sec:maths} establishes the formulae to calculate the optimal number of aggregated frames which, on average, saves transmission times. Finally, Section~\ref{sec:conclusions} concludes the paper.

\section{Related Work}
\label{sec:related_work}

Most of the published works study the performance of  IEEE 802.11  under different conditions, such as~\cite{Bianchi-00,XR-CommLett-02,TF-CommLett-06}, to name a few. Other works closer to ours are~\cite{TDSK-WiCOM-12}, where a packet aggregation scheme is proposed to maximize network performance, and~\cite{KWT-MICC-09}, where the frame delay is studied under stochastic terms. 

Other works have studied the savings achievable when using 802.11 aggregation for packets from one application (or different applications) to the same destination~\cite{SS10}. In~\cite{SRA17}, a central-controlled aggregation mechanism is proposed that limits the maximum aggregation size in applications with real-time constraints that share the medium with background traffic. This mechanism enables the provision of a low latency service for applications with real-time constraints and maximum throughput for the other applications. As an alternative, the authors also consider   permanently setting a low value for the maximum A-MPDU size. An adaptive machine learning-based approach to adjusting the maximum length of A-MSDU per user is presented in~\cite{CTR20}. In this context, the authors of~\cite{SRF+21} propose an algorithm for dynamically adjusting the maximum size of aggregated frames in 802.11 WLANs, which allows a network administrator to find an optimal balance between performance and latency in these networks.

As far as we know, we are the first to mathematically study when delaying frame transmissions saves transmission time, and how much time it saves, across all IEEE 802.11 protocols.

\section{Stochastic Transmission Time}
\label{sec:stochastic}

In this section, we present a stochastic interpretation of the transmission time that we then use in our calculations.


\begin{figure*}[h]
\centering
\scalebox{1}{
\begin{tikzpicture}[node distance = 5pt]

  \makeatletter

  \tikzstyle{label}=[
  font=\footnotesize,
  ]

  \tikzstyle{box}=[
  draw=black,
  rectangle,
  minimum height=40pt,
  text centered]

  \node (PLCPp) [box, minimum width=40pt, text width=40pt] {PLCP\\preamble + header};
  \node (BACKOFF) [box, dashed, minimum width=30pt, left = of PLCPp] {backoff};
  \node (DIFS) [box, minimum width=30pt, left = of BACKOFF] {DIFS};
  \node (MPDU_MAC) [box, minimum width=30pt, right = of PLCPp, text width=30pt] {MAC\\header};
  \node (MPDUd) [box, dashed, minimum width=70pt, right = of MPDU_MAC] {Data};
  \node (MPDUcrc) [box, minimum width=10pt, right = of MPDUd] {CRC};
  \node (SIFS) [box, minimum width=30pt, right = of MPDUcrc] {SIFS};
  \node (PLCPp2) [box, minimum width=40pt, text width=40pt, right = of SIFS] {PLCP\\preamble + header};
  \node (ACK) [box, minimum width=30pt, text width=30pt, right = of PLCPp2] {ACK};

%

  \draw [<->] ($(PLCPp.north west) + (0, 5pt) $) -- ($(PLCPp.north east) + (0, 5pt) $) node [label, above, midway] { $t_{pr}$ };
  \draw [<->] ($(BACKOFF.north west) + (0, 5pt) $) -- ($(BACKOFF.north east) + (0, 5pt) $) node [label, above, midway] { $t_{\text{\em backoff}}$ };
  \draw [<->] ($(DIFS.north west) + (0, 5pt) $) -- ($(DIFS.north east) + (0, 5pt) $) node [label, above, midway] { $t_{DIFS}$ };
  \draw [<->] ($(MPDU_MAC.north west) + (0, 5pt) $) -- ($(MPDUcrc.north east) + (0, 5pt) $) node [label, above, midway] { $t_{tr}$ };
   \draw [<->] ($(MPDU_MAC.south west) - (0, 5pt) $) -- ($(MPDU_MAC.south east) - (0, 5pt) $) node [label, below, midway] { $t_{MAC}$ };
   \draw [<->] ($(MPDUd.south west) - (0, 5pt) $) -- ($(MPDUd.south east) - (0, 5pt) $) node [label, below, midway] { $t_{data}$ };
 	\draw [<->] ($(MPDUcrc.south west) - (0, 5pt) $) -- ($(MPDUcrc.south east) - (0, 5pt) $) node [label, below, midway] { $t_{crc}$ };
	\draw [<->] ($(SIFS.north west) + (0, 5pt) $) -- ($(SIFS.north east) + (0, 5pt) $) node [label, above, midway] { $t_{SIFS}$ };
	\draw [<->] ($(PLCPp2.north west) + (0, 5pt) $) -- ($(PLCPp2.north east) + (0, 5pt) $) node [label, above, midway] { $t_{pr}$ };
	\draw [<->] ($(ACK.north west) + (0, 5pt) $) -- ($(ACK.north east) + (0, 5pt) $) node [label, above, midway] { $t_{ack}$ };
\end{tikzpicture}
}
	\caption{Single frame transmission under IEEE 802.11 (not scaled).}
	\label{fig:wifi_frame}
\end{figure*}
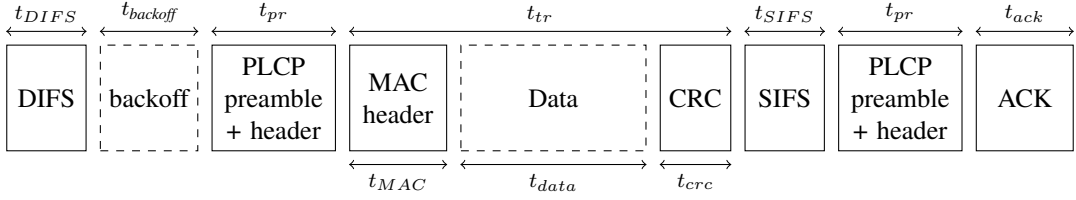

Figure~\ref{fig:wifi_frame} shows the timing diagram involved in a successful  single frame transmission under IEEE 802.11 (adapted from~\cite{D-PIMRC-08}). Some of these times are overhead, such as DIFS and SIFS intervals, ACK transmission, and the Physical Layer Convergence Protocol (PLCP) preamble and header that precede each frame. These values depend on the characteristics of the physical layer. The MAC header  and CRC  transmission times are also dependent on the physical layer bit rate. 

The time boxes depicted with a dashed line in Figure~\ref{fig:wifi_frame} indicate that there is some randomness and therefore, we cannot define them as a concrete value: $t_{\text{\em backoff}}$ refers to the backoff time, which is equal to a discrete random value uniformly distributed in the interval $\{0, \text{CW}\}$, multiplied by $\text{SLOT}$ units of time; and $t_{data}$ refers to the data payload transmission, which is equal to a discrete random value that expresses the size of the data (in bytes) multiplied by the time required to transmit a byte, considering the underlying physical characteristics.

Let us first denote $t_{\text{\em backoff}}$ in terms of random variables. Let $X\sim \mathcal{U}_{\{0, 1, ..., \text{CW}\}}$ be a discrete random variable following a discrete uniform distribution $\mathcal{U}$ in the interval $\{0, 1, ..., CW\}$. Therefore, $Y=\text{SLOT} \cdot X \sim \mathcal{U}_{\{0, \text{SLOT}, \dots, \text{SLOT}\cdot \text{CW}\}}$. Let the time  $t^i_{\text{\em backoff}}$ for a frame $i$ be denoted as $Y_i$. Although this model does not closely match reality (since when the medium is congested and collisions are generated, the backoff times are redefined to $\{0, 1, \ldots, 2\text{CW}\}$), our final goal is to compare the times of the  standard model with those of our proposal (see Figure~\ref{fig:sketches}), where there is no probability of collision since we have grouped all the sources in a single buffer, which cannot collide. Therefore, to assume that there are no collisions in a system without frame aggregation is to give it an additional advantage  when comparing its performance with our proposal of frame aggregation.

Now, let us denote $t_{data}$ similarly to $t_{\text{\em backoff}}$. Let $P$ be a discrete random variable that describes the size distribution (in bytes) of the data payload of the frames. Let $Z= \frac{8}{br} P$ be the dependent variable that represents the time required to transmit a data payload, where $br$ is the physical layer bit rate, in units of bits per second. Let the transmission time $t^i_{data}$ for a frame $i$ be denoted as $Z_i$ seconds.

Putting everything together, the total time $tx_i$ for a successful transmission of a frame $i$ under IEEE 802.11 is defined as 
%
$tx_i = \gamma + Y_i + Z_i,$
%
 where $\gamma=t_{DIFS} + 2t_{pr} + t_{MAC} + t_{crc}+t_{SIFS}+t_{ack}$ is a real constant whose value depends on the underlying physical characteristics defined in IEEE 802.11.
\section{On the Optimal Number of Frames Aggregated}
\label{sec:maths}

This section first describes our proposal to aggregate frames intuitively and then details the assumptions and derivation of the mathematical formulae to calculate the number of aggregated frames that, on average, will allow us to improve the transmission time. Finally, we present a series of numerical experiments and provide our final observations.

\subsection{Description of the Intuitive Idea}

Figure~\ref{fig:standard_model} depicts the standard model of a node that receives $N$ transmission sources, each with a distribution rate of $\lambda_i, 1 \leq i \leq N$. Each destination frame $f_i$ is queued, assigned a backoff time $t^{f_i}_{\text{backoff}}$, and sent accordingly when the backoff time expires and the medium is free, in accordance with IEEE 802.11.

Our proposed frame-aggregation model is outlined in Figure~\ref{fig:proposed_model}. Unlike the standard model (without frame aggregation), the destination frames are queued together. Upon the arrival of $k$ frames, they are aggregated and sent as a single multicast frame $g$ following the IEEE 802.11 specification (i.e., a backoff time $t^{g}_{\text{backoff}}$ is calculated accordingly and the frame $g$ is sent when the backoff time expires). Thus, through multicast transmission, downlink efficiency is increased by aggregating packets from different applications and various origins and destinations. Security mechanisms can be put in place to prevent group members from reading data that is not intended for them. 

Our aggregation proposal, via multicast frames, is implementable in all IEEE 802.11 standards, thus improving the efficiency achieved regardless of the chosen standard. The use of multicast frames also has some drawbacks. For instance, it is more difficult to provide QoS support in the case of multicast traffic. In addition, multicast transmissions on 802.11 networks face serious reliability and scalability problems, due to the lack of acknowledgement messages and retransmissions. For this reason, the 802.11 standard recommends the use of basic data rates, which makes the channel busy for longer periods of time, affecting other network services. To solve these problems, the IEEE 802.11aa amendment introduced a set of retransmission policies to provide robust audio and video services in multicast mode, although the concrete way to use these features is left to the implementer~\cite{802.11-standard-12}.


\tikzset{font=\small}
\begin{figure}
\centering
\subfigure[Standard model (without frame aggregation)]{
\begin{tikzpicture}[stack/.style={rectangle split, minimum width=10mm,rectangle split parts=#1,draw, anchor=center}]
	\node[stack=3] (lambda1) { };
	\node[above of = lambda1,node distance=15mm] (tlambda1) { };
	\draw[->] (tlambda1) -- node[right] {$\lambda_1$} (lambda1);
	
	\node[stack=3,right of=lambda1,node distance=15mm] (lambda2) { };
	\node[above of = lambda2,node distance=15mm] (tlambda2) { };
	\draw[->] (tlambda2) -- node[right] {$\lambda_2$} (lambda2);
	
	\node[right of=lambda2,node distance=15mm] (lambda){ };
	\node[above of = lambda,node distance=15mm] (tlambda) {\bf Sources};
	\node[below of=tlambda,node distance=10mm] (ttlambda){\bf \ldots};
	
	\node[stack=3,right of=lambda,node distance=15mm] (lambdaN1) { };
	\node[above of = lambdaN1,node distance=15mm] (tlambdaN1) { };
	\draw[->] (tlambdaN1) -- node[right] {$\lambda_{N-1}$} (lambdaN1);

	\node[stack=3,right of=lambdaN1,node distance=15mm] (lambdaN) { };
	\node[above of = lambdaN,node distance=15mm] (tlambdaN) { };
	\draw[->] (tlambdaN) -- node[right] {$\lambda_{N}$} (lambdaN);
	
	\node[draw,circle,below of=lambda,node distance=15mm] (tx) {\bf Tx};
	\node[below of=tx,node distance=10mm] (txx) { };
	
	\draw[->] (lambda1.south) -- (tx);
	\draw[->] (lambda2.south) -- (tx);
	\draw[->] (lambdaN1.south) -- (tx);
	\draw[->] (lambdaN.south) -- (tx);
	\draw[->,line width=0.5mm] (tx) -- (txx);
	
	\draw[dashed] ($(lambda1.west) - (3mm,2mm)$) -- node[below,text width=25mm,align=center] {\scriptsize \em Backoff countdown position} ($(lambdaN.east) + (3mm,-2mm)$);
\end{tikzpicture}
\label{fig:standard_model}
}
\subfigure[Our proposed frame-aggregation model]{
\begin{tikzpicture}[decoration={brace,mirror},stack/.style={rectangle split, minimum width=10mm,rectangle split parts=#1,draw, anchor=center}]
	\node[ ] (lambda1) {};
	\node[above of = lambda1,node distance=5mm] (tlambda1) {$\lambda_1$};
	
	\node[right of=lambda1,node distance=15mm] (lambda2) { };
	\node[above of = lambda2,node distance=10mm] (tlambda2) {$\lambda_2$};
	
	\node[yshift=10pt,right of=lambda2,node distance=15mm] (lambda){\bf \ldots};
	\node[above of = lambda,node distance=5mm] (tlambda) {\bf Sources};
	
	\node[,right of=lambda2,node distance=40mm] (lambdaN1) { };
	\node[above of = lambdaN1,node distance=10mm] (tlambdaN1) {$\lambda_{N-1}$};

	\node[right of=lambdaN1,node distance=15mm] (lambdaN) { };
	\node[above of = lambdaN,node distance=10mm] (tlambdaN) {$\lambda_{N}$};
	
	\node[stack=4,below of=lambda2,node distance=10mm] (buffer) { };
	\node[left of=buffer] (buffert) {\em Buffer};
	
	\draw [decorate,line width=0.5pt] 
  ([xshift=3pt]buffer.south east) -- node[right,yshift=.5mm] (sizek) {size $k$} ([yshift=-25pt,xshift=3pt]buffer.north east);
  
	\draw[->] (tlambda1) -- ([xshift=-3mm]buffer.north);
	\draw[->] (tlambda2) -- (buffer.north);
	\draw[->] (tlambdaN1) -- ([xshift=3mm]buffer.north);
	\draw[->] (tlambdaN) -- ([xshift=4mm]buffer.north);
	
	\node[stack=3,below of=lambdaN1,node distance=10mm] (buffer2) {};
	
	\draw[dashed] ($(buffer2.west) - (5mm,2mm)$) -- node[right,text width=25mm,align=center] {\scriptsize \em Backoff countdown position} ($(buffer2.east) + (15mm,-2mm)$);
	
	\node[draw,circle,below of=buffer2,node distance=15mm] (tx) {\bf Tx};
	\node[below of=tx,node distance=10mm] (txx) { };
	
	\draw[->] (buffer2.south) -- (tx);
	\draw[->,line width=0.5mm] (tx) -- (txx);
	
	  \draw[dashed,->]  (sizek.north) .. controls +(0,1) and +(0,1).. (buffer2.north);

\end{tikzpicture}
\label{fig:proposed_model}
}
\label{fig:sketches}
\caption{Sketches of the transmission model of a node.}
\end{figure}
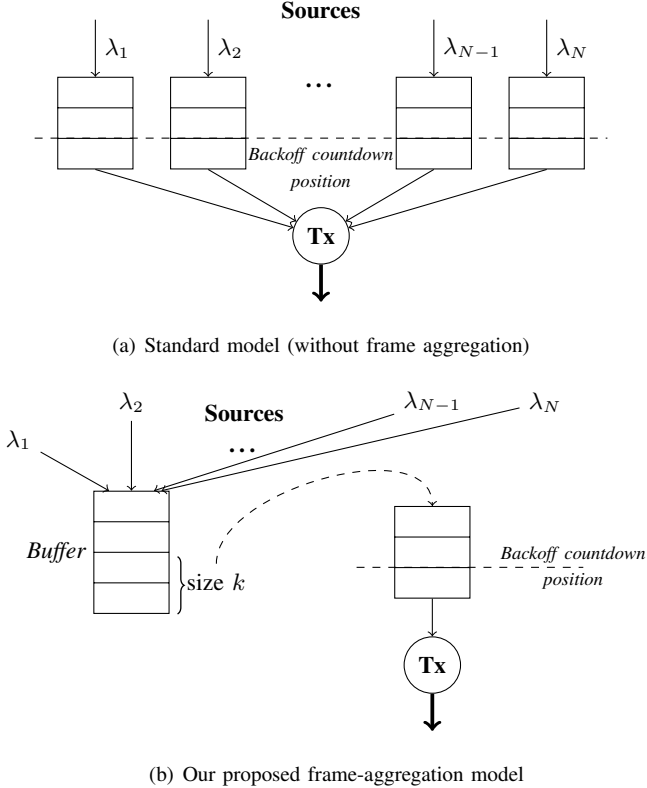

\subsection{Calculating the Optimal $k$ Value}

Initially, we start from a series of sources that send frames to a transmission service. Let us focus on one of these sources, $A$, with a distribution rate $\lambda_i$. We will assume that all the distribution rates follow a Poisson distribution. Therefore, the distribution rate resulting from adding all of them is also a Poisson distribution whose rate is the sum of the rates, denoted as $\lambda$. We assume a payload size for each source, which follows a distribution (usually exponential), denoted as $P$. In any event, the payload size distribution is left open, because we are primarily interested in the mean. 

Our scenario is the following:

\begin{enumerate}
	\item The frames enter and wait for a buffer of size $k$. Since they enter with a Poisson distribution, the waiting time for the arrival of $k$ packets follows an Erlang distribution of mean $\dfrac{k}{\lambda}$. Then, the mean waiting time of the $j$th frame in the buffer will be $\dfrac{k - j}{\lambda}$, on average $\text{Er}(k)=\dfrac{k - 1}{2\lambda}$.
	\item With a time that we will consider negligible, an aggregated frame $f$ of size $k\text{E}[P]$ is built that will wait for a mean backoff time $t^f_{\text{backoff}}$, where $P$ is the random variable that determines the payload size of the frames that will be transmitted.	
	\item The aggregated frame will then move to a queuing system with a Poisson input rate $\lambda_A (k) = \lambda/k$. The service time for each frame will be $\dfrac{1}{\mu(k)} = \dfrac{8\cdot k\text{E}[P]}{br} + \gamma + t^f_{\text{backoff}}$, where $\mu(k)$ is the system throughput, measured as bits per second (bps). We consider that the average time of stay of the frame in the system is $W (k)$. As it is a $M / G / 1$ queue, hence $W (k) = \dfrac{(\lambda_A(k))^2(\sigma(K))^2 + (\rho(k))^2}{2\lambda_A(k)(1 - \rho(k))}$, where $\rho(k) = \dfrac{\lambda_A(k)}{\mu(k)}$. Another formulation (Pollaczek-Khintchine~\cite{Pollaczek1930,Khintchine1932}) can be $W (k) = \dfrac{\lambda_A(k)(\mu(k))^{-2}}{2\left( 1-\rho(k)\right)}$.
	\item The mean time of the frame in the system will be $F (k) = \text{Er}(k) + \dfrac{1}{\mu(k)} + W (k)$. 
\end{enumerate}

Therefore, we can evaluate the gains of our proposed model (shown in  Figure~\ref{fig:proposed_model}) by evaluating $G(k) = F(k) - F(1)$. When $k \in \mathbb{Z}_{>1}$ and  $G(k) > 0$, then the aggregation of $k$ frames causes, on average, that the delay in receiving all frames in the destination is less than sent individually.

\subsection{Numerical Experiments}

To appreciate under what circumstances our proposal on frame aggregation will be most beneficial, we need to calculate $G(k)$ with network traffics that have different $\lambda$ distribution rates (measured in pps) and varying the value of $k$.

\begin{figure}
	\centering
	\includegraphics[width=.99\columnwidth]{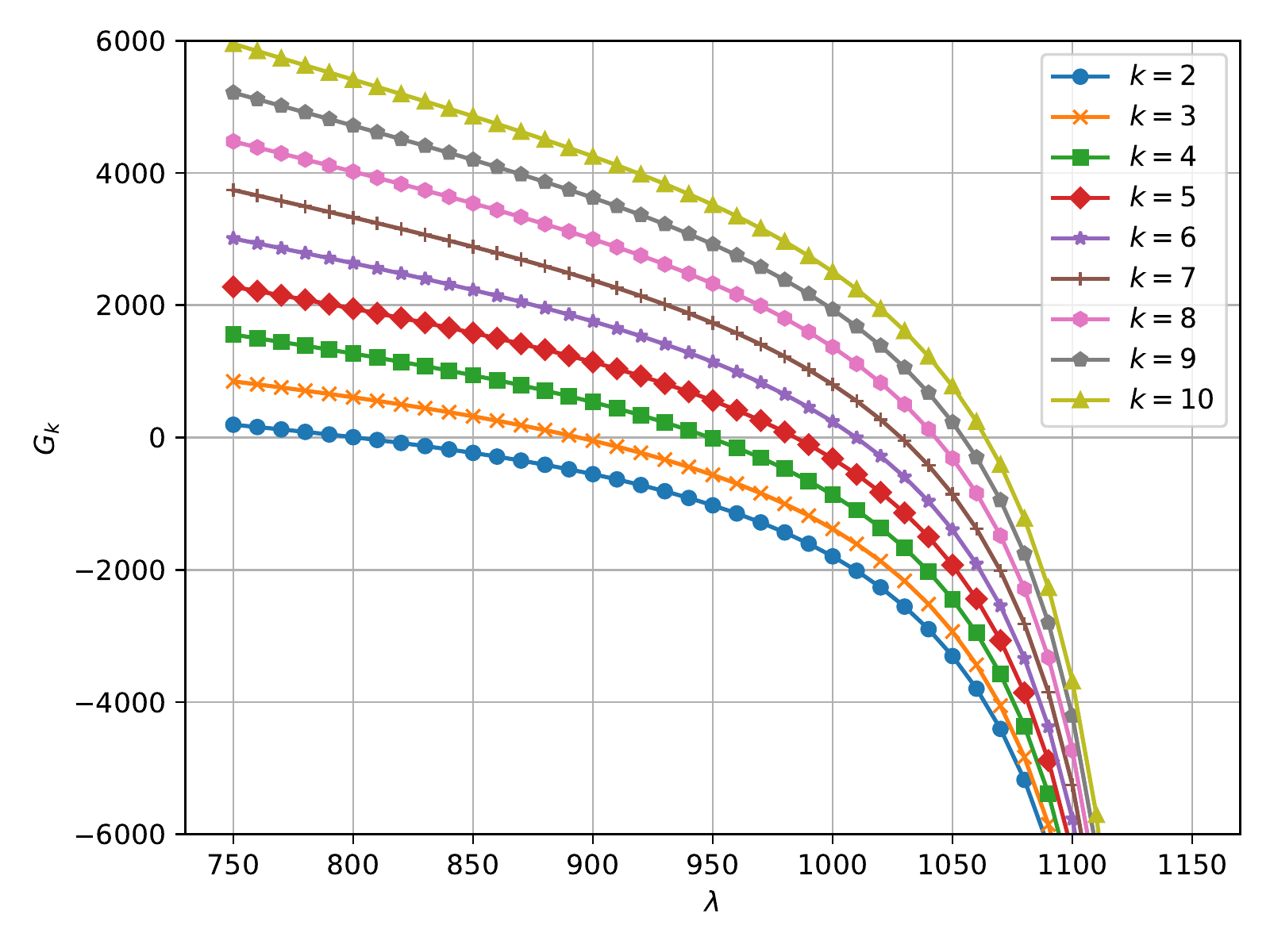}
	\caption{Values of $G(k)$ in an IEEE 802.11b transmission of 11Mbps rate, when $k \in \{2, 3, ..., 10\}$ ($CW=16, t_{\text{backoff}}=20\mu s, t_{DIFS}= 50\mu s, t_{pr}=96 \mu s,$ and $E[P]=800$ bits).}
	\label{fig:Gk}
\end{figure}

Figure~\ref{fig:Gk} shows the value of $G(k)$ and $\lambda$ calculated for a WiFi transmission under IEEE 802.11b with a rate of 11Mbps, for different values of $k (k \in {2, 3, ..., 10})$. The other IEEE 802.11b parameters remain the same for all experiments ($CW=16, t_{\text{backoff}}=20\mu s, t_{DIFS}= 50\mu s, t_{pr}=96 µs$ and $E[P]=800$ bits). To better understand what the values of $\lambda$ mean, remember that they are proportional to the input rate values following the expression $\lambda E[P]$, measured in bps. We can observe that, regardless of the number of frames aggregated, $G(k)$  takes positive values up to a certain value of $\lambda$. A positive value indicates that a delay in the transmission will appear if we aggregate frames. From that value of $\lambda$, $G(k)$ takes negative values, which indicates that the aggregation no longer causes delay. Furthermore, the graph also indicates that if the value of $\lambda$ continues to increase, the values of $G(k)$ tend to $-\infty$. This means that if no aggregation is done, the transmission becomes impossible because there is a situation of unacceptable congestion.

\begin{figure}
	\centering
	\includegraphics[width=.99\columnwidth]{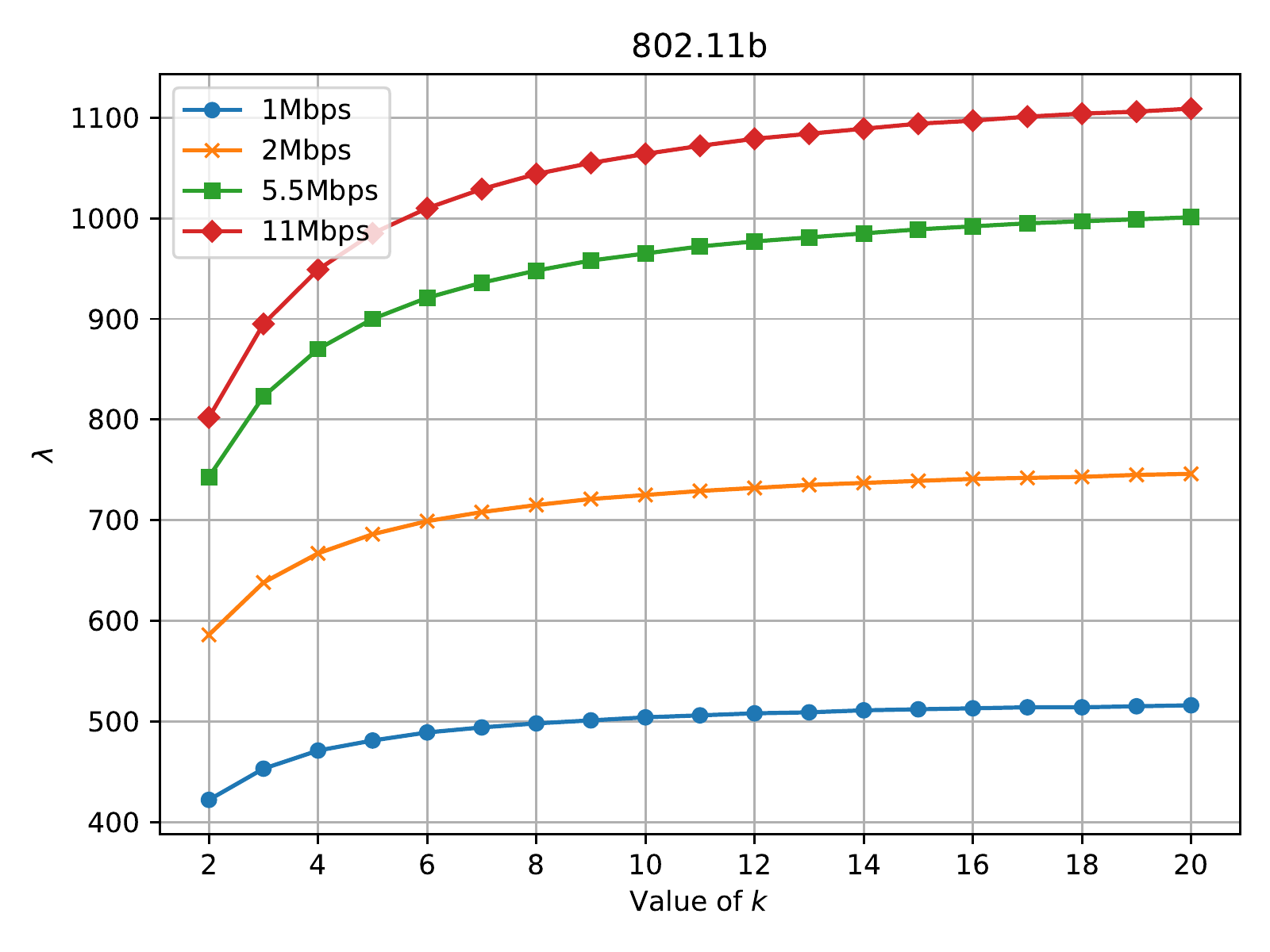}
	\caption{Value of $\lambda$ that makes $G(k), k \in \{2, 3, ..., 20\},$ a negative value for different rates of an IEEE 802.11b transmission ($CW=16, t_{\text{backoff}}=20\mu s, t_{DIFS}= 50\mu s, t_{pr}=96\mu s$ and $E[P]=800$ bits).}
	\label{fig:IEEE802.11b}
\end{figure}

\begin{figure}
	\centering
	\includegraphics[width=.99\columnwidth]{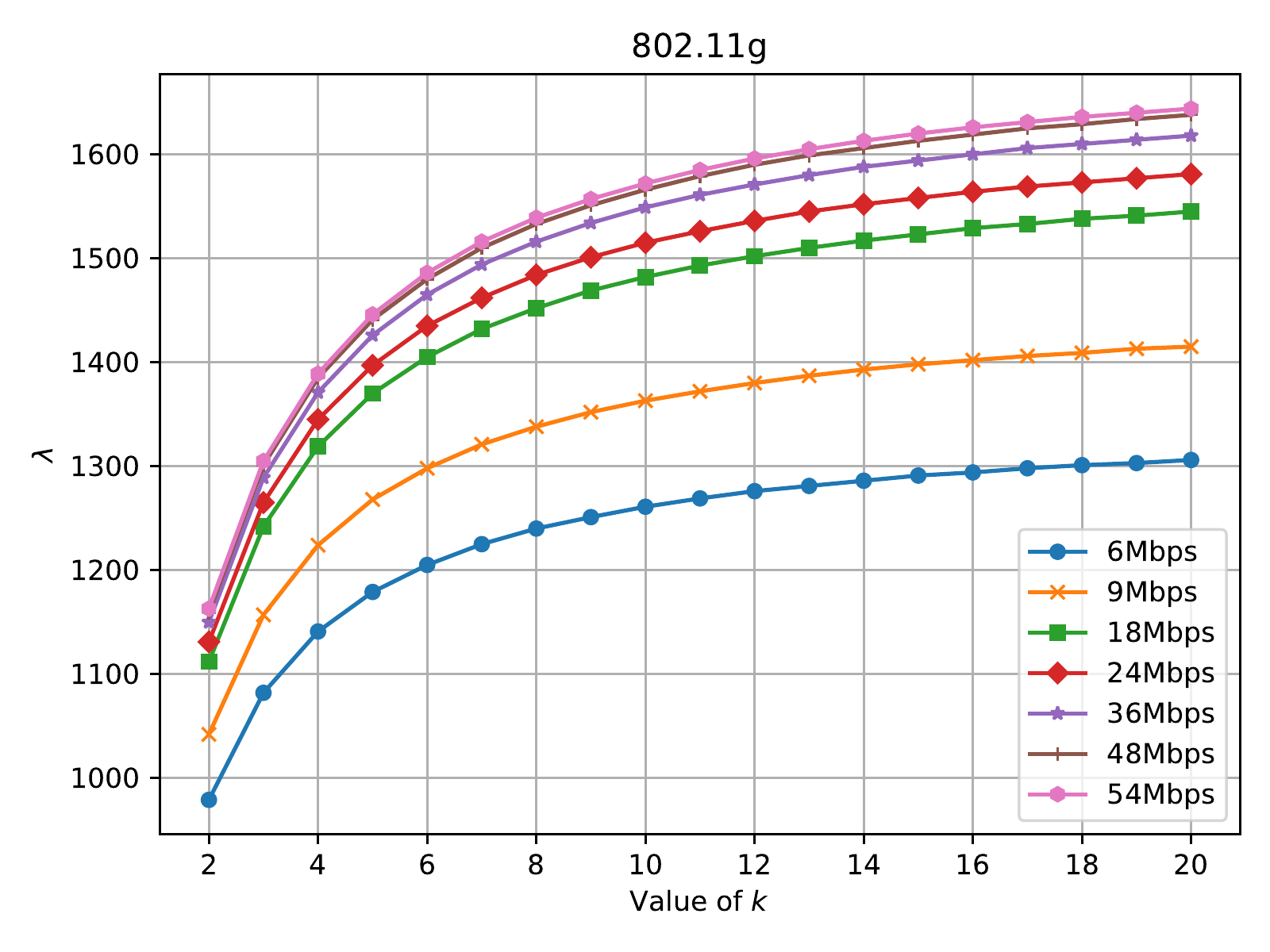}
	\caption{Value of $\lambda$ that makes $G(k), k \in \{2, 3, ..., 20\},$ a negative value for different rates of an IEEE 802.11g transmission ($CW=16, t_{\text{backoff}}=20\mu s, t_{DIFS}= 28\mu s, t_{pr}=22.1\mu s$ and $E[P]=800$ bits).}
	\label{fig:IEEE802.11g}
\end{figure}

Figures~\ref{fig:IEEE802.11b} and~\ref{fig:IEEE802.11g} show the value of $\lambda$ that makes $G(k), k \in {2, 3, ..., 20},$ become negative (and thus the aggregation no longer causes delay, but attenuates it) for the transmission rates specified by the 802.11b and 802.11g WiFi standards, respectively. The other parameters used for these numerical experiments are shown in the title of each figure. We can see that when the value of $k$ increases, the value of $\lambda$ also increases. This finding is logical, because for values greater than $k$, the delay increases unless the input rate is higher, in which case the aggregation decreases the delays. 

{\em Final Remarks.} In summary, we have found that the delays caused by the aggregation of frames are compensated by the decrease in transmission congestion, which is a recommended practice to allow a better use of the radio medium in WiFi. To conclude, frame aggregation not only improves congestion situations, but also improves transmission delays. The mathematical expression proposed in this work is useful to discriminate when to use frame aggregation and save transmission time, on average. This expression can be used as a step prior to transmission, but the final decision of the value $k$ chosen must be adapted to each case.

\section{Conclusions}
\label{sec:conclusions}

We have described the transmission time of a frame under the IEEE 802.11 protocol in a stochastic way and then we have obtained a formula to calculate the number of frames that can be aggregated to save transmission time, on average. Our numerical experiments demonstrate that WiFi frame aggregation improves the efficiency under certain IEEE 802.11 traffic conditions saving transmission time and reducing transmission congestion, and can be even proposed in those standards that do not have  an aggregation mechanism defined. This formula can be used as a configuration criterion for a traffic regulator node (for instance, a router) with frame aggregation  capability, to decide when frame aggregation should occur.
%


\bibliographystyle{IEEEtran}
\bibliography{biblio}

\end{document}